# Synthesizing 3D computed tomography from MRI or CBCT using 2.5D deep neural networks

Satoshi Kondo[1][0000-0002-4941-4920], Satoshi Kasai[2] and Kousuke Hirasawa[3]

[1] Muroran Institute of Technology, Hokkaido, Japan
[2] Niigata University of Health and Welfare, Niigata, Japan
[3] Konica Minolta, Inc., Osaka, Japan
kondo@muroran-it.ac.jp

**Abstract.** Deep learning techniques, particularly convolutional neural networks (CNNs), have gained traction for synthetic computed tomography (sCT) generation from Magnetic resonance imaging (MRI), Cone-beam computed tomography (CBCT) and PET. In this report, we introduce a method to synthesize CT from MRI or CBCT. Our method is based on multi-slice (2.5D) CNNs. 2.5D CNNs offer distinct advantages over 3D CNNs when dealing with volumetric data. In the experiments, we evaluate the performance of our method for two tasks, MRI-to-sCT and CBCT-to-sCT generation. Target organs for both tasks are brain and pelvis.

## 1      Introduction

Radiation therapy (RT) is a critical cancer treatment that often requires computed tomography (CT) for accurate dose calculations. Magnetic resonance imaging (MRI) provides superior soft tissue contrast, but lacks the electron density data of CT for dose calculations. Combining the two modalities presents challenges, including misregistration errors.

MRI-only RT has emerged to address these challenges, reduce ionizing radiation exposure, and improve patient comfort. However, the generation of synthetic CT images from MRI (sCT) remains challenging due to the lack of direct correlation between nuclear magnetic properties and electron density.

Deep learning (DL) techniques, particularly convolutional neural networks (CNNs), have gained traction for sCT generation from MRI, Cone-beam CT (CBCT) and PET [1].

In this report, we introduce a method to synthesize CT from MRI or CBCT. Our method is based on multi-slice (2.5D) CNNs. 2.5D CNNs offer distinct advantages over 3D CNNs when dealing with volumetric data. These benefits stem from a thoughtful compromise between computational efficiency and capturing relevant spatial context. In the experiments, we evaluate the performance of our method for two tasks, MRI-to-sCT and CBCT-to-sCT generation. Target organs for both tasks are brain and pelvis.

## 2  Proposed Method

Our base method is same for both tasks and both organs. We use encoder-decoder type deep neural networks for converting MRI or CBCT images to synthetic CT (sCT) images.

Figure 1 shows an overview of our method. Although the input images are 3D volumes, we use a 2D deep neural network model with multi-slice inputs (2.5D CNNs). 2.5D CNNs offer distinct advantages over 3D CNNs when dealing with volumetric data. These benefits stem from a thoughtful compromise between computational efficiency and capturing relevant spatial context. Reasons why 2.5D CNNs are favored in many cases include reduced computational complexity, memory efficiency, leveraging anisotropic resolution, multi-planar analysis, contextual information, and overcoming class imbalance.

In our model, $N$ consecutive slices in an input volume are processed to produce one slice in a sCT volume. The input slices are along transverse plane. The consecutive slices are processed as an $N$ channel 2D image in our model. In training phase, $N$ slices are randomly selected $M$ times from each volume in the training dataset in each epoch. In inference phase, each volume is processed in slice-by-slice way and each slice in sCT volume is produced.

We use L1 error between predicted sCT slices and ground truth CT slices as the loss function.

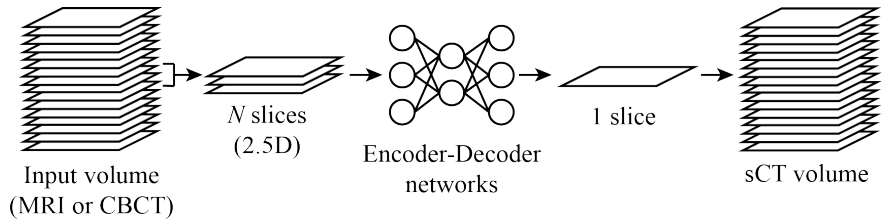

**Fig. 1.** Overview of our method.

## 3  Experiments

### 3.1  Dataset

Data was acquired for radiotherapy treatments in the radiotherapy departments of UMC Utrecht, UMC Groningen, and Radboud Nijmegen) [2]. The numbers of data are summarized in Table 1. Each data includes source image (MRI for the MRI-to-sCT task and CBCT for the CBCT-to-sCT task), ground truth (CT) and mask.

We divide each dataset to training and validation data. The numbers of training and validation data are 162 and 18 in each dataset, respectively.

Table 1. Datasets

| Task | Organ | Number of data |
|---|---|---|
| MRI-to-sCT | Brain | 180 |
| | Pelvis | 180 |
| CBCT-to-sCT | Brain | 180 |
| | Pelvis | 180 |

### 3.2 Experimental conditions

We used U-Net [3] as the basic segmentation network and replaced its encoder part as EfficientNet [4]. We conducted hyper-parameter tuning. The hyper-parameters include the encoder size, the number of slices, the initial learning rate. As the results of hyper-parameter tuning, we selected EfficientNet-B7 as the encoder, 3 as the number slices. The initial learning rates were selected as $1\times10^{-3}$, $5\times10^{-4}$, $1\times10^{-4}$, and $5\times10^{-5}$ for task-1 brain, task-1 pelvis, task-2 brain, and task-2 pelvis, respectively.

The optimizer was AdamW [5] and the learning rate was decreased at every epoch with cosine annealing. The number of epochs was 100, and We used the model with the lowest loss value for the validation data as the final model.

As pre-processing, histogram normalization was performed for MRI volumes. No data augmentations were performed.

### 3.3 Experimental results

Table 2 shows the summary of the experimental results. We show two metrics; PSNR and Mean Absolute Error (MAE). These are the differences between sCT and ground truth CT. As for the tasks, it cannot be seen big differences between MRI-to-sCT and CBCT-to-sCT.

Figures 2, 3, 4 and 5 show examples of experimental results. In each figure, (a) shows an input slice (MRI or CBCT), (b) shows the corresponding slice of sCT, and (c) shows the corresponding slice of ground truth (CT).

Table 2. Experimental results for validation dataset.

| Task | Organ | PSNR (dB) ↑ | Mean Absolute Error (HU) ↓ |
|---|---|---|---|
| MRI-to-sCT | Brain | 27.06 | 77.93 |
| | Pelvis | 28.51 | 64.26 |
| CBCT-to-sCT | Brain | 27.38 | 81.44 |
| | Pelvis | 28.12 | 68.07 |

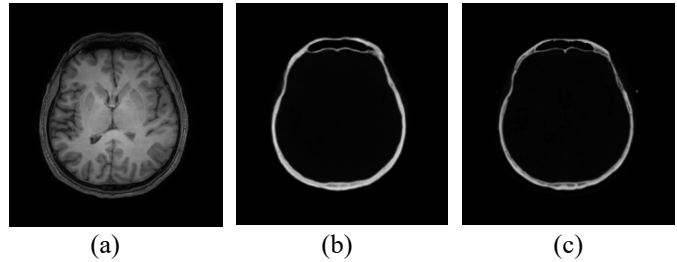

(a) (b) (c)

**Fig. 2.** Examples of experimental results in MRI-to-SCT / Brain. (a) MRI (input). (b) sCT (output). (c) CT (ground truth).

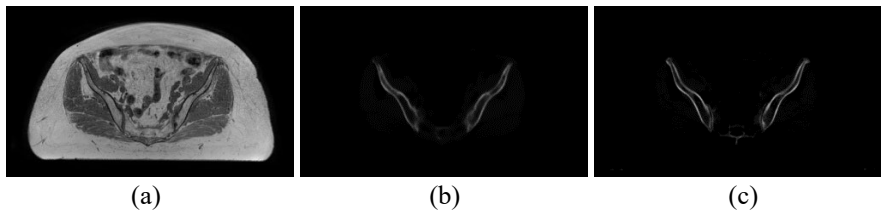

(a) (b) (c)

**Fig. 3.** Examples of experimental results in MRI-to-SCT / Pelvis. (a) MRI (input). (b) sCT (output). (c) CT (ground truth).

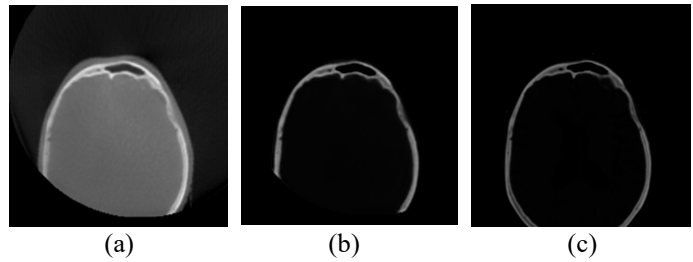

(a) (b) (c)

**Fig. 4.** Examples of experimental results in CBCT-to-SCT / Brain. (a) CBCT (input). (b) sCT (out-put). (c) CT (ground truth).

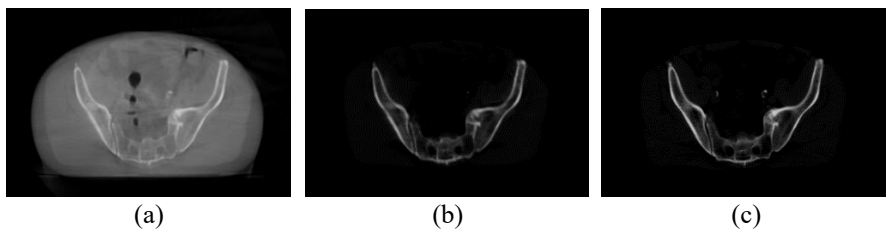

(a) (b) (c)

**Fig. 5.** Examples of experimental results in CBCT-to-SCT / Pelvis. (a) CBCT (input). (b) sCT (out-put). (c) CT (ground truth).

We also evaluated our method in SynthRAD2023 challenge site [6]. In preliminary test task, the algorithm run six cases on the grand challenge platform and the system gives MAE, PSNR and SSIM metrics for each case. Table 3 shows the summary of the preliminary test task.

Table 3. Experimental results in preliminary test task. Each value shows "mean±standard deviation" for six cases.

| Task | PSNR (dB) ↑ | Mean Absolute Error (HU) ↓ | SSIM ↑ |
|---|---|---|---|
| MRI-to-sCT | 23.99±1.91 | 85.23±20.35 | 0.82±0.04 |
| CBCT-to-sCT | 23.64±1.26 | 91.59±18.55 | 0.77±0.06 |

## 4  Conclusions

In this report, we introduced a method to synthesize CT from MRI or CBCT. Our method is based on multi-slice (2.5D) CNNs. In the experiments, we evaluatde the performance of our method for two tasks, MRI-to-sCT and CBCT-to-sCT generation. Target organs for both tasks are brain and pelvis. From the experimental results, big differences in performance between MRI-to-sCT and CBCT-to-sCT were not observed. As for the organs, the results for pelvis were slightly better than the results for brain.